\documentclass[conference,letterpaper]{IEEEtran}

\usepackage[T1]{fontenc}
\usepackage[utf8]{inputenc}
\usepackage{cite}
\usepackage{amsmath,amssymb,amsfonts}
\usepackage{algorithmic}
\usepackage{graphicx}
\usepackage{textcomp}
\usepackage{xcolor}
\usepackage{listings}
\usepackage{booktabs}
\usepackage{url}
\usepackage{balance}
\usepackage{syntax}

\usepackage[bookmarks=false]{hyperref}
\hypersetup{
  colorlinks=true,
  linkcolor=black,
  citecolor=black,
  urlcolor=blue
}

\definecolor{dkgreen}{rgb}{0,0.6,0}
\definecolor{ltblue}{rgb}{0,0.4,0.4}
\definecolor{dkviolet}{rgb}{0.3,0,0.5}
\definecolor{dkblue}{rgb}{0,0.1,0.7}
\definecolor{dkred}{rgb}{0.7,0,0}

\lstdefinelanguage{MutAlg}{
  morekeywords={},
  sensitive=true,
  morestring=[b]",
  literate={+}{{{\color{blue}+}}}1
           {*}{{{\color{blue}*}}}1
           {-}{{{\color{blue}-}}}1,
}

\lstdefinelanguage{Rust}{
  morekeywords={
    as,break,const,continue,crate,else,enum,extern,false,fn,for,if,impl,in,let,
    loop,match,mod,move,mut,pub,ref,return,self,Self,static,struct,super,trait,
    true,type,unsafe,use,where,while
  },
  sensitive=true,
  morecomment=[l]{//},
  morecomment=[s]{/*}{*/},
  morestring=[b]",
}

\lstdefinestyle{haskellstyle}{
  language=Haskell,
  mathescape=true,
  basicstyle=\ttfamily\small,
  columns=[l]flexible,
  identifierstyle={\ttfamily\color{black}},
  keywordstyle={\ttfamily\color{dkviolet}},
  keywordstyle=[2]{\ttfamily\color{dkblue}},
  keywordstyle=[3]{\ttfamily\color{ltblue}},
  stringstyle={\ttfamily\color{dkred}},
  commentstyle={\ttfamily\upshape\color{dkgreen}},
  morekeywords=[2]{case,of,let,in,where,if,then,else,do},
  morekeywords=[3]{IO,Int,Integer,Bool,Char,Maybe,Either},
}

\lstdefinestyle{ruststyle}{
  language=Rust,
  basicstyle=\ttfamily\small,
  columns=[l]flexible,
  identifierstyle={\ttfamily\color{black}},
  keywordstyle={\ttfamily\color{dkviolet}},
  stringstyle={\ttfamily\color{dkred}},
  commentstyle={\ttfamily\upshape\color{dkgreen}},
}

\lstdefinelanguage{Json}{
  morestring=[b]",
  stringstyle=\color{dkred},
  sensitive=true,
  keywords={true,false,null},
  keywordstyle=\color{dkblue},
}

\lstdefinestyle{jsonstyle}{
  language=Json,
  basicstyle=\ttfamily\small,
  columns=[l]flexible,
  identifierstyle={\ttfamily\color{black}},
  keywordstyle=\color{dkblue},
  stringstyle=\color{dkred},
}

\lstdefinelanguage{Diff}{
  morecomment=[f][\color{dkblue}]{@@},
  morecomment=[f][\color{dkgreen}]{+},
  morecomment=[f][\color{dkred}]{-},
}

\lstdefinestyle{diffstyle}{
  language=Diff,
  basicstyle=\ttfamily\small,
  columns=[l]flexible,
}

\lstdefinestyle{bnf}{
  basicstyle=\ttfamily\small,
  columns=[l]flexible,
  keepspaces=true,
  showstringspaces=false,
  keywordstyle={\ttfamily\color{dkviolet}},
  morekeywords={expr,tag,mutant,mutation},
  literate={::=}{{{\color{dkblue}::=}}}3
           {|}{{{\color{dkblue}|}}}1
           {+}{{{\color{dkred}+}}}1
           {*}{{{\color{dkred}*}}}1,
}

\newcommand{\tool}[0]{{Marauder}}

\begin{document}

\title{A Declarative Framework for Hand-Crafted\\Mutation Analysis and Management}

\author{\IEEEauthorblockN{Alperen Keles}
\IEEEauthorblockA{University of Maryland, College Park \\ akeles@umd.edu}}

\maketitle

\begin{abstract}
Hand-crafted mutants are increasingly used to evaluate fuzzing and
property-based testing tools, but current tooling is fragmented and often
forces trade-offs between readability, mutation preservation, and execution
cost. We present a declarative framework for hand-crafted mutation analysis and management.
First, we characterize five mutation representations: comment-based,
preprocessor-based, patch-based, match-and-replace, and in-AST
mutations. Second, we define a mutation algebra that supports selective execution, tag-based
expansion, and higher-order combinations of mutants. Third, we describe a
lossless conversion pipeline that maps mutation representations through a
common intermediate form, including a strategy for extracting and
normalizing in-AST mutations. We implement these ideas in \tool{}, a
prototype system for injecting, activating, resetting, and composing
hand-crafted mutants across representations. This framework clarifies the
design space of hand-crafted mutation systems and provides a practical foundation
for more efficient and expressive mutation experiments.
\end{abstract}

\begin{IEEEkeywords}
mutation testing, mutation analysis, property-based testing, declarative
languages, hand-crafted mutants
\end{IEEEkeywords}

%% ============================================================
\section{Introduction}
\label{sec:introduction}

Mutation testing is a cornerstone of evaluating the effectiveness of testing~\cite{jia2011analysis}.
In a static test suite, mutation testing reveals untested behavior by breaking existing
code, which a sufficiently comprehensive testing suite must catch, to detect weaknesses
in the testing coverage. In dynamic testing tools such as model checkers, fuzzers,
property-based testing tools, mutation testing is used to empirically evaluate the
strength of the method~\cite{hritcu2013testingnoninterference}. Over the years, many mutation testing tools across different
languages and ecosystems have proliferated the testing ecosystem~\cite{shi2023etna}. These tools
have, by a large margin, focused on \emph{automating} mutations. For instance,
one could implement a very simple mutation tool that goes through any mathematical
operators in the code and switch them with some alternatives, or change comparisons,
or change number or string literals. Production-grade tools, of course, are much
more nuanced on their strategies of mutation, as well as the \emph{level} of mutation
itself. While some tools such as mutants.rs~\cite{cargomutants}, mutmut~\cite{mutmut} or StrykerJS~\cite{strykerjs}
apply source-level mutations before compilation or interpretation, PIT~\cite{pit} mutates
the JVM bytecode generated by the compiler, Major~\cite{JustSK2011} is a compiler plugin that traverses
the AST and embeds the mutations in the generated bytecode, and Mull~\cite{denisov2018mull} creates mutations
in-memory at the level of LLVM bitcode, injecting them to the program to be run without
recompilation of C++ programs. The level of the mutations present a core trade-off in the design of a mutation
testing tool. Source-level mutations are easily presented to the user as diffs
while lower level changes can be more efficient or require less user intervention.

In this paper, we focus on an unusual scenario in the mutation testing literature,
\emph{hand-crafted mutations}. Whereas automated mutation testing relies on automatically
synthesized mutations via syntactic and semantic analysis of the underlying programs,
hand-crafted mutations rely on expert judgement.
These mutations are typically mined out of historical precedent, previously fixed
bugs on established data structures, algorithms or software products~\cite{shi2023etna}.
Compared to automated mutation analysis, the tools for hand-crafted mutation analysis have
garnered very little attention. Throughout the rest of this paper, we start by building
a framework of source-level hand-crafted mutation syntaxes, followed by an algorithm
that can convert each mutation to another without any loss of information, and an
implementation of a new hand-crafted mutation injection and analysis tool, \tool{}.

The contributions of the paper are as follows:

\begin{enumerate}
  \item We present five different types of hand-crafted mutation systems
  and outline the advantages and disadvantages of each system.
  \item We develop an algorithm for lossless conversions between each mutation system
  we presented.
  \item We present \tool{}, an implementation of the techniques we presented.
\end{enumerate}

%% ============================================================
\section{Background and Motivation}
\label{sec:background}

Why do we care about manual mutations if they aren't the major focus of the literature?
The reason is that hand-crafted mutations are increasingly relied upon, relative to automated
mutations, in the evaluation of random testing tools in fuzz testing and property-based testing
literature. In~\cite{klees2018evaluatingfuzztesting}, the authors present a series of
recommendations for realistic and robust evaluation of fuzz testing tools, Magma\cite{hazimeh2020magma}
is a ground-truth fuzzing benchmark that relies on the said advice, which we
concisely outline based on the systematization in FixReverter\cite{zhang2022fixreverter}:

\begin{enumerate}
  \item the benchmark should use relevant, real-world target programs;
  \item programs should contain realistic, relevant bugs
  (e.g., memory corruption/crash bugs);
  \item bugs should be triggerable in a way that clearly indicates
  when a particular bug is found, to avoid problems with deduplication;
  \item the benchmark should defend against overfitting.
\end{enumerate}

The property-based testing literature follows a similar strategy for evaluating PBT tools.
In~\cite{hritcu2013testingnoninterference}, the authors test an information flow control abstract
machine by manually injecting bugs in the machine instructions. In~\cite{hughes2020howtospecifyit},
the author presents a series of techniques for writing properties and uses hand-crafted
mutations in data structures for evaluation, these works are brought together in ETNA\cite{shi2023etna},
which builds an evaluation and analysis platform for PBT. ETNA consists of five different
workloads: Binary Search Tree, Red-Black Tree, Simply-Typed Lambda Calculus, System F,
and IFC, with more than 100 hand-crafted mutations in total. These mutations are managed
using a comment-based syntax; we present a bug in the insertion operation of a binary search tree in Haskell,
taken from the ETNA paper:

\begin{lstlisting}[style=haskellstyle]
insert k v E = T E k v E
insert k v (T l k v r)
  {-! -}
  | k < k = T (insert k v l) k v r
  | k > k = T l k v (insert k v r)
  | otherwise = T l k v r
  {-!! insert_becomes_singleton -}
  {-!
  = T E k v E
  -}
\end{lstlisting}

This bug, \texttt{insert_becomes_singleton}, changes the tree to hold only
one key-value pair at a time. The comment syntax is inspired by QuickChickTool\cite{lampropoulos2018quickchickbook},
a command-line tool for interacting with QuickChick\cite{quickchickrepo}, a descendant of QuickCheck\cite{claessen2000quickcheck}
in Rocq Theorem Prover. The start of a variation is marked by
\texttt{\{-! -\}} that can have multiple alternative mutations, each marked by a header
\texttt{\{-!! <mutation> -\}} followed by a body \texttt{\{-! <code> -\}}.

This design is fundamentally flawed in one aspect: it is not \emph{mutation-preserving}.
Activating the last mutation removes the mutation boundary, so the system cannot
keep track of the mutation structure in the code. ETNA manages this by keeping
the original project intact, copying it to a temporary directory, and applying
the mutation there instead of changing anything inline.

There's another crucial drawback of the comment-based mutation syntax,
it requires recompiling the project for each mutation. In the automated mutation-analysis
landscape, we see a shift from source-rewriting in mutmut~\cite{mutmut} in Python or StrykerJS~\cite{strykerjs}
in JavaScript where recompilations are relatively low-cost, to more bytecode~\cite{pit} or
bitcode~\cite{denisov2018mull} oriented mutations in Java and C++ as well as manipulating the AST via
metaprogramming in Rust. The trade-off in the situation is very clear, as the cost of
compilation is higher, tools move from recompilation to reinterpretation of the same code
with slight runtime costs instead of high compilation costs. We see a similar situation
in ETNA mutations, where almost all of the existing languages in the benchmark have
high compilation costs, so many times the bottleneck to running experiments is the
compilation instead of running the PBT tools themselves.

These problems indicate a trade-off space in the hand-crafted mutation landscape as well
as opportunities for development in the current methods. In the next section, we build
a framework/taxonomy of possible mutation systems for hand-crafted mutations. We lay
out a series of requirements, discuss the advantages and disadvantages of each option
and sketch out designs for each option.

\section{An Exploration of Hand-Crafted Mutations}
\label{sec:theory}

What are the properties we expect from a hand-crafted mutation system?
As we already expressed in the previous section, we would like the system
to be \emph{mutation-preserving}, activating a mutation should not break
our ability to further analyze the program. Furthermore, there's a trade-off
between the presentability of the mutation versus the efficiency based on
the level of the mutation. We add another dimension to the argument,
\emph{language-awareness}, denoting the level of syntactic detail the mutation tool
possesses on the mutated program. Comment-based mutations we introduced earlier
are a middle ground between AST-level language integration of in-AST mutations we'll
introduce in Section~\ref{sec:functional} and the language agnostic
mutations we'll introduce shortly.

%% ============================================================
\subsection{Comment-Based Mutations}
\label{sec:comment}

Comment-based mutations we mentioned in the previous section
had two main issues: mutation preservation and compilation cost. During
our implementation, we encountered another one, the syntactic marker
we use (\texttt{<comment-begin>! <comment-end>}) may already be reserved in the
language. For instance, Rust reserves (\texttt{/*!}) for documentation purposes
to the point that incorrect usage results in a compilation error. That's why in addition
to parameterizing (\texttt{<comment-begin>}) and (\texttt{<comment-end>}), we
also started to parameterize (\texttt{<mutation-marker>}), for which we are using
(\texttt{|}) instead of (\texttt{!}) for Rust.

We solved the problem of mutation preservation by adding an end marker after
the last mutant ((\texttt{<comment-begin> !<comment-end>})). This allows us to
set, unset, reset any mutations in the program without breaking the capability
to analyze further, or any further mutation injections. Below, we present
an example of a comment-based mutation system for the same bug we had in the previous section
in Rust:

\begin{lstlisting}[style=ruststyle]
match t {
  E => T(E, k, v, E),
  T(l, k2, v2, r) => {
    /*| insert */
    if k < k2 {
      T(insert(k, v, *l), k2, v2, r)
    } else if k2 < k {
      T(l, k2, v2, insert(k, v, *r))
    } else {
      T(l, k2, v, r)
    }
    /*|| insert_1 */
    /*| T(E, k, v, E) */
    /* |*/
  },
}
\end{lstlisting}

%% ============================================================
\subsection{Preprocessor-Based Mutations}
\label{sec:preprocessor}

We can have a preprocessor-based mutation system, where we have a series of preprocessor
flags for each mutation, and the mutation is activated by setting the corresponding
flag. This system is mutation-preserving with an additional cost of managing active
mutations as the code itself is unaware of the activation; the activation flag
is set at compilation time. This system is completely language-agnostic
as it relies on the preprocessor and no language-specific details are encoded in
the mutation system itself. Below is an example of a preprocessor-based mutation system:

\begin{lstlisting}[language=C]
#if defined(M_INSERT_1) /* variation=insert */
...1
#elif defined(M_INSERT_2)
...2
#elif defined(M_INSERT_3)
...3
#else
...4
#endif

\end{lstlisting}

%% ============================================================
\subsection{Patch Mutations}
\label{sec:patch}

The third system we present is based on patch files kept in a separate patches
directory. In adition to the base source file, we usea manifest file that stores bundle metadata
(including source path and variation tags), and one unified diff file per variant.
Activating a mutation corresponds to applying the selected variant patch to the base fragment.
Below is an example of a patch-based mutation block:

\begin{lstlisting}[style=diffstyle]
diff --git a/calc.rs b/calc.rs
--- a/calc.rs
+++ b/calc.rs
@@ -2,1 +2,1 @@
-    a + b
+    a - b
\end{lstlisting}

%% ============================================================
\subsection{Match and Replace Mutations}
\label{sec:match-and-replace}

The last language-agnostic mutation system we present is based on a match-and-replace
representation encoded as JSON. Each mutation block records a scoped region in a file,
the base pattern to match, and replacement snippets for each mutant variant. This keeps
mutation data structured and language-agnostic while still preserving enough positional
information to reconstruct the mutation structure in the source code. We define a match and replace mutation as follows:

\begin{lstlisting}[style=jsonstyle]
{
  "name": "add",
  "scope": "calc.rs:2",
  "match": "a + b",
  "variants": [
    { "name": "add_1", 
      "replacement": "a - b" },
    { "name": "add_2",
      "replacement": "a * b" }
  ]
}
\end{lstlisting}

The change is applied by matching the recorded pattern inside the stored scope and
replacing it with the selected variant replacement.

%% ============================================================
\subsection{In-AST Mutations}
\label{sec:functional}

The last mutation system we present is a language-aware system that is based on runtime-activated
mutations embedded in the AST. Similar to the preprocessor-based system, the code itself is
unaware of the active mutations, so the mutations are managed by a runtime system that
keeps track of the active mutations and activates them at runtime. This system has the
unique advantage compared to the other systems in that it removes the need for recompilation
between multiple mutation passes. However, the mutations are very much intertwined with the code itself,
the code is much less presentable and harder to analyze. Below, we present an example of an in-AST mutation
system for the same bug we had in the previous section:

\begin{lstlisting}[style=ruststyle]
T(l, k2, v2, r) => {
  match () {
    _ if mutation_active("insert_1") => {
      T(E, k, v, E)
    }
    _ => { // base
      if k < k2 { 
        T(insert(k, v, *l), k2, v2, r)
      } else if k2 < k {
        T(l, k2, v2, insert(k, v, *r))
      } else {
        T(l, k2, v, r)
      }
    }
  }
}
\end{lstlisting}

%% ============================================================
\subsection{Mutation Algebra}
\label{sec:algebra}

In an automated mutation system, mutations are generated online by the mutation tool,
so the natural behavior of the system is to test all generated mutants in some order.
In a hand-crafted system, we might want to test a subset of the mutants, test combinations
of them at the same time, or even test them in a specific order. For instance, we might
want to start from "easy" mutants that are more likely to be caught by the testing suite, and then move on to
harder mutants that are less likely to be caught. How should we let the user specify these combinations
and orders? We defined a simple algebra of mutations for expressing these combinations:

\begin{lstlisting}[style=bnf]
expr ::= expr + expr
       | expr * expr
       | +tag
       | *tag
       | mutant
       | mutation
\end{lstlisting}

Tags we introduced in the algebra are optional annotations for the mutations, for instance we can
mark a mutation as "easy" or "hard" to indicate the difficulty of the mutation:

\begin{lstlisting}[style=haskellstyle]
{-!! insert_1 [easy, small] -}
\end{lstlisting}

When evaluating a mutation expression, we interpret the unary operators on the tags as expansions,
\texttt{+tag} is expanded to $m_1 + m_2 + \dots + m_n$ where $m_i$ are all mutations with tag $tag$,
and \texttt{*tag} is expanded to $m_1 * m_2 * \dots * m_n$ where $m_i$ are all mutations with tag $tag$.
A \texttt{mutation} is expanded to the sum of its mutants, so \texttt{insert} is expanded to
\texttt{insert_1 + insert_2 + insert_3}. The plus (+) operator is interpreted as testing the
mutants in sequence, so $m_1 + m_2$ means testing $m_1$ first and then testing $m_2$. The multiplication (*)
means testing the mutants in parallel, so $m_1 * m_2$ means activating $m_1$ and $m_2$ at the same time
and testing the combination. So if we have a group of mutants we would like to test at the same time,
we can mark them with a tag and use the multiplication operator to test them together.

For complex expressions, we compute the expansion of the expressions and then turn the formula into
sum-of-products form, so we are only left with a sequence of products, which are mutants to activate
at a time. If a product contains more than one mutant from the same mutation, we throw an error
as mutants for the same mutation are mutually exclusive and cannot be activated at the same time.

%% ============================================================
\section{Mutation Conversion}
\label{sec:conversion}

The mutation systems we presented in the previous section all have advantages and disadvantages,
all but the in-AST mutations have compilation costs, the comment-based system is easy to analyze
and read, and the preprocessor and patch systems are completely language-agnostic,
match and replace is convenient for multiple simultaneous mutations. It would be ideal
if we could convert between these systems without any loss of information, so we can have the best of all worlds.

We built a series of bridges between each system and the comment-based system for enabling any pairwise conversion.
The conceptual model of a program is a series of code and mutations interleaved together, each mutation with a base code and some set of mutants.
Once we can construct this data format from any one of the mutation systems, we can then render it
back to any of the other systems. For the comment and preprocessing based systems, we simply parse
the code and extract the mutations based on the syntax. For the patch-based system, we parse
the metadata block and unified diff hunks to recover mutation locations and replacements.
For the match and replace system, the mutations are already in a structured JSON format.

The challenge arises from the in-AST mutation system due to two distinct issues. The first is extraction:
how do we extract the mutations from the code when the mutations are embedded in the AST and not explicitly
marked in the code? We can solve this via embedding the mutations with some reserved markers such as the
\texttt{match ()} call that would be very unusual in a Rust program. The larger issue is the mismatch of
the mutation model between the in-AST system and the other systems. In the in-AST system, each mutation
must be a valid syntactic unit in the language, so there are mutations in other systems that cannot be
directly represented in the in-AST system. Below, we present a simple example:

\begin{lstlisting}[style=ruststyle]
foo(
/*| foo */
  0)
/*|| foo_1 */
  1)
/* |*/
\end{lstlisting}

Both of these mutants are valid in the comment-based system, the base program is \texttt{foo(0)},
the mutated program is \texttt{foo(1)}. However, the mutation breaks through the syntax of the program,
so we cannot translate this mutation in a straightforward way to the in-AST system.

We solve this by finding the smallest syntactic unit that contains all of the mutants, meaning that
for any activated mutant, the node we end up with is a valid syntactic unit in the language. So we transform
the above mutation to the following:

\begin{lstlisting}[style=ruststyle]
/*| foo */
  foo(0)
/*|| foo_1 */
  foo(1)
/* |*/
\end{lstlisting}

Which can be easily translated to the in-AST system as follows:

\begin{lstlisting}[style=ruststyle]
match () {
  _ if mutation_active("foo_1") => {
    foo(1)
  }
  _ => {
    foo(0)
  }
}
\end{lstlisting}

\begin{table*}[t!]
\centering
\caption{Comparison of Compilation and Execution Times between Comment-Based and In-AST Mutations}
\label{tab:build-exec-model}
\resizebox{\textwidth}{!}{
\begin{tabular}{lrrrrrrrrrrr}
\toprule
\textbf{Workload} & \textbf{n} & \textbf{Comment Cold (s)} & \textbf{Comment Warm (s)} & \textbf{Comment Exec (s)} & \textbf{Comment Total (s)} & \textbf{In-AST Cold (s)} & \textbf{In-AST Exec (s)} & \textbf{In-AST Total (s)} & \textbf{Compilation Speedup} & \textbf{Exec Slowdown} \\
\midrule
BST  & 8  & 20.17 & 2.47 & 0.04 & 37.51  & 20.35 & 0.05 & 20.40 & 1.84$\times$ & 1.30$\times$ \\
RBT  & 13 & 18.34 & 1.70 & 2.67 & 41.39  & 19.75 & 2.99 & 22.74 & 1.82$\times$ & 1.12$\times$ \\
STLC & 10 & 19.81 & 1.19 & 36.64 & 67.17 & 20.28 & 39.29 & 59.57 & 1.13$\times$ & 1.07$\times$ \\
\midrule
\textbf{Total} & \textbf{31} & \textbf{58.32} & \textbf{5.36} & \textbf{39.35} & \textbf{146.07} & \textbf{60.38} & \textbf{42.34} & \textbf{102.72} & \textbf{1.42$\times$} & \textbf{1.08$\times$} \\
\bottomrule
\end{tabular}
}
\end{table*}

%% ============================================================
\section{\textsc{Marauder}: Implementation}
\label{sec:marauder}

We implemented all the mutation systems we presented in the previous section, as well as the conversion algorithm
for Rust ASTs as a Rust library and command-line tool, \tool{}. The tool currently works for any language
in the language-agnostic systems, for Haskell, Rocq, Racket, OCaml, Rust, and Python for the comment-based system,
and Rust for the in-AST system. The public API of the tool, exposed as a command-line interface, is
as follows:

\noindent \textbf{list}: list all the mutations in the project, along with their tags and locations.\\
\textbf{set / unset}: set/unset a mutant as active, the mutant is specified by its name.\\
\textbf{reset}: unset all mutations.\\
\textbf{test}: run a given command for all evaluated mutant sets for a mutation expression.\\
\textbf{convert}: convert the mutation system of the project to another system, the target system is specified as an argument.

Additionally, we implemented an import feature from \texttt{cargo-mutants} to our
mutation system, which allows the users to start from an automated mutation system
and gradually pick out the mutants they want to keep and convert them to hand-crafted mutations for further analysis.

% Our implementation of the in-AST mutation system, at the cost of a small runtime overhead, has allowed
% us to speed up some of the experiments in the ETNA benchmarking repository immensely. We present
% the results of our evaluation in Section~\ref{sec:eval}.

\begin{figure}[t!]
  \includegraphics[width=\linewidth]{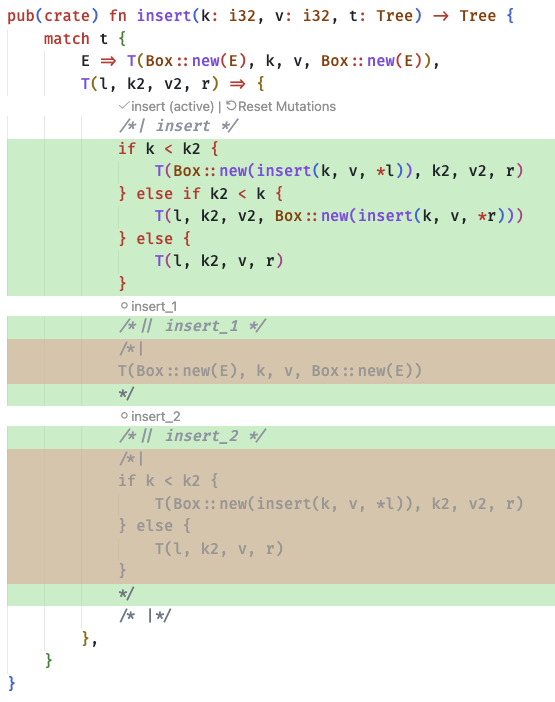}
  \caption{IDE Plugin for Mutation Management}
  \label{fig:ide}
\end{figure}

We also built a \tool{} interface in the ETNA IDE plugin for Visual Studio Code that allows users
to interact with the mutations in the code, such as activating, deactivating, resetting, and converting mutations.
Figure~\ref{fig:ide} shows a screenshot of the plugin in action.

\section{Evaluation} \label{sec:eval}

In our evaluation, we compare the performance of the in-AST mutation system with the comment-based mutation system on the ETNA benchmark.
We run the full benchmarks for Binary Search Tree, Red-Black Tree, and Simply-Typed Lambda Calculus workloads in Rust with both mutation
systems, measure cold and warm compilation times as well as execution times for each mutant, construct a model of the total time taken
in each scenario, and compare the results. The results are summarized in Table~\ref{tab:build-exec-model}.

\subsection{Performance: In-AST vs.\ Comment-Based Mutations}

In our Rust workloads, compilation can dominate runtime by several orders of
magnitude, leading per-mutant recompilation to be the bottleneck. We ran
all three workloads with both mutation systems using deterministic random seeds
to remove generator randomness as a confounder. The in-AST system showed a significant
reduction in total time taken compared to the comment-based system ranging from
1.13$\times$ to 1.84$\times$ speedup in compilation time, while keeping
the execution slowdown to a minimum, ranging from 1.07$\times$ to 1.30$\times$. To
understand the effects of the compilation and execution times on the total time taken,
we built a simple model as follows:

\noindent $T_{comment} = compile_{cold} + (n - 1) \times compile_{warm} + execution$\\
$T_{in-ast} = compile_{cold} + execution$

We assume one cold compilation at the beginning, followed by $n-1$ warm compilations for the comment-based system for
each mutant in the workload, and one cold compilation for the in-AST system. The execution times are added to the total
time for both systems. The slowdown in execution times is positively correlated with the total time of the workload.
This is presumably because the runtime overhead from runtime mutation management is less significant
as the rest of the computation becomes more expensive.

%% ============================================================
\section{Conclusions and Future Work}
\label{sec:conclusion}

We presented a framework for hand-crafted mutation analysis, which we see as an increasingly
important part of the mutation testing ecosystem in the future. Automated mutation testing is
a powerful tool for evaluation of testing suites, especially with the rise of
LLM-driven mutations~\cite{tip2025llmorpheus, dakhel2024mutap}, the need to store mutations,
analyze, manipulate, and manage them is crucial. We plan to expand on potential
mutation systems we have not yet explored with a more comprehensive analysis of the trade-offs and design space
as well as build on the in-AST mutation system to be able to support more languages as well as
prove the correctness of the conversion algorithm beyond the currently limited testing
we have done in our implementation.

%% ============================================================
%% References
%% ============================================================

\balance
\bibliographystyle{IEEEtran}
\bibliography{references}

\end{document}